\documentclass[preprint]{aastex}
\input{epsf}

\newcommand{\gsim}{\mathrel{\hbox{\rlap{\hbox{\lower4pt\hbox{$\sim$}}}\hbox{$>$}}}}
\newcommand{\lsim}{\mathrel{\hbox{\rlap{\hbox{\lower4pt\hbox{$\sim$}}}\hbox{$<$}}}}

\shortauthors{Hoyle et al.} 
\shorttitle{Two-Dimensional Topology of the SDSS}

\begin{document}

\title{Two-Dimensional Topology of the Sloan Digital Sky Survey}

\author{Fiona Hoyle$^1$, Michael S. Vogeley$^1$, J. Richard Gott
III$^2$, Michael Blanton$^3$, Max Tegmark$^4$, David H. Weinberg$^5$, Neta A. Bahcall $^2$, Jon Brinkmann$^6$ \& Donald G. York$^7$  \\ 
1. Department of Physics, Drexel University, 3141
Chestnut Street, Philadelphia, PA 19104 \\ 2. Princeton University
Observatory, Peyton Hall, Princeton, NJ 08544\\ 3. Department of Physics, New York University, 4 Washington Place, New York, NY 10003 \\ 4. Department of Physics, University of Pennsylvania, Philadelphia, PA, 19101 \\ 5. Department of Astronomy, Ohio State University, 140 W. 18th Ave., Columbus, OH 43210 \\ 
6. Apache Point Observatory, P. O. Box 59, Sunspot, NM 88349-0059 \\ 
7. The University of Chicago, Astronomy \& Astrophysics Center, 5640 S. Ellis Ave., Chicago, IL, 60637 \\
\email{hoyle@venus.physics.drexel.edu, vogeley@drexel.edu,
jrg@astro.princeton.edu} }

\begin{abstract}

We present the topology of a volume-limited sample of 11,884 galaxies,
selected from an apparent-magnitude limited sample of over 100,000
galaxies observed as part of the Sloan Digital Sky Survey (SDSS). The
data currently cover three main regions on the sky: one in the Galactic
north and one in the south, both at zero degrees declination, and one
area in the north at higher declination. Each of these areas covers a
wide range of survey longitude but a narrow range of survey latitude,
allowing the two dimensional genus to be measured.

The genus curves of the SDSS sub-samples are similar, after
appropriately normalizing these measurements for the different areas.
We sum the genus curves from the three areas to obtain the total genus
curve of the SDSS. The total curve has a shape similar to the genus
curve derived from mock catalogs drawn from the Hubble Volume
$\Lambda$CDM simulation and is similar to that of a Gaussian random
field.  Likewise, comparison with the genus of the 2dFGRS, after
normalization for the difference in area, reveals remarkable
similarity in the topology of these samples.  

We test for the effects of galaxy type segregation by splitting the
SDSS data into thirds, based on the $u^*-r^*$ colors of the galaxies,
and measure the genus of the reddest and bluest sub-samples. This
red/blue split in $u^*-r^*$ is essentially a split by morphology
(Strateva et al. 2001).  We find that the genus curve for the
reddest galaxies exhibits a ``meatball'' shift of the topology
-- reflecting the concentration of red galaxies in high density regions --
compared to the bluest galaxies and the full sample, in agreement with
predictions from simulations.

\end{abstract}


\section{Introduction}

Quantitative measurements of the large-scale structure of the galaxy
distribution provide important constraints for models of structure
formation. In particular, the topology of large-scale structure is a
useful test of the initial conditions of the Universe.  Inflation
predicts that the seeds for structure formation should derive from a
Gaussian random phase distribution (Bardeen, Steinhardt \& Turner
1983). The topology of the large-scale structure is invariant during
the linear growth of structure. Thus, after appropriate smoothing, the
topology of the present galaxy distribution can be related to that of
the initial density field. The topology can be quantified using
statistical methods, such as the genus statistic (Gott et
al. 1986). This statistic allows a test of the random phase
hypothesis, because any deviation of the measured topology on large
smoothing scales might be evidence for non-Gaussian initial
conditions. On relatively smaller scales, the topology of the smoothed
galaxy density quantifies the degree of non-linear evolution and/or
biasing of galaxy formation with respect to the mass density at the
present epoch. The genus statistic quantifies the topology of the
smoothed galaxy distribution by computing the integrated curvature of
iso-density contours at different density thresholds.  This statistic
is complementary to statistical measures such as the power spectrum or
correlation function, which quantify only the second moments of the
distribution. By focusing on the connectedness of structure as a
function of filling factor, the genus statistic also complements the
1-point probability distribution of smoothed density fields, which has
a very different sensitivity to non-linearity, bias, and primordial
non-Gaussianity.  Analyses of early SDSS data include measurement of
the power spectrum (Szalay et al. 2002; Tegmark et al. 2002; Dodelson
et al. 2002), correlation function (Zehavi et al. 2002; Connolly et
al. 2002; Infante et al. 2002) and higher order moments (Szapudi et
al. 2002).

The genus statistic has been measured from a variety of galaxy
surveys.  Gott et al. (1989) applied the 3D genus statistic to small
samples of galaxies. It has since been applied to subsequently larger
surveys; the 3D genus of the SSRS was measured by Park, Gott \& da
Costa (1992), Moore et al. (1992) applied it to the QDOT Survey,
Rhoads et al. (1994) analyzed Abell Clusters, Vogeley et al. (1994)
analyzed the CfA survey, Canavezes et al. (1998) analyzed the PSCz
Survey, Protogeros \& Weinberg (1997) and Springel et al. (1998)
analyzed the 1.2-Jy IRAS Survey. Hikage et al. (in preparation)
estimate the 3D genus for an early sample of galaxies from the SDSS.
Melott (1987) was the first to mention the possibility of studying the
2D topology, i.e. the genus of 1-dimensional contours of smoothed
surface density in a thin slice redshift survey.  This idea was
further developed by Melott et al. (1989). The 2D genus method has
been applied to the Lick Galaxy Catalog (Coles \& Plionis 1991), the
Abell and ACO cluster catalogs (Plionis, Valdarnini \& Coles 1992),
the CfA Survey (Park et al. 1992) and the LCRS survey (Colley
1997). Hoyle, Vogeley \& Gott (2002, HVG02) estimate the
two-dimensional genus for the 2dF Galaxy Redshift. Some evidence was
found for departure from Gaussianity that would be expected as a
result of non-linear gravitational evolution and/or biasing of
galaxies as compared to the mass but on the whole, the results were
consistent with the universe having Gaussian initial conditions.

In this paper, we estimate the genus of an initial sample of SDSS
galaxies. This is the first time that a sample from the SDSS is large
enough that measurements of the two-dimensional genus can be
made. Because the SDSS includes not only medium-resolution
spectroscopy but also five-band digital photometry for every object in
the redshift sample, we can split the sample as a function of color,
which is a good indicator of morphology. For the first time, we
examine the two-dimensional genus of the reddest and bluest galaxies
to see if genus is a function of galaxy type.

We describe the SDSS galaxy survey in Section \ref{sec:data} and summarize the
method for measuring the genus in section \ref{sec:genus}. We present
our results in section \ref{sec:res} and we draw conclusions in
section \ref{sec:conc}.

\section{The Data Set and the Simulation}
\label{sec:data}
\subsection{The Sloan Digital Sky Survey}
\label{sec:survey}

\begin{figure} 
\begin{centering}
\begin{tabular}{c}
{\epsfxsize=8truecm \epsfysize=8truecm \epsfbox[0 170 650 600]{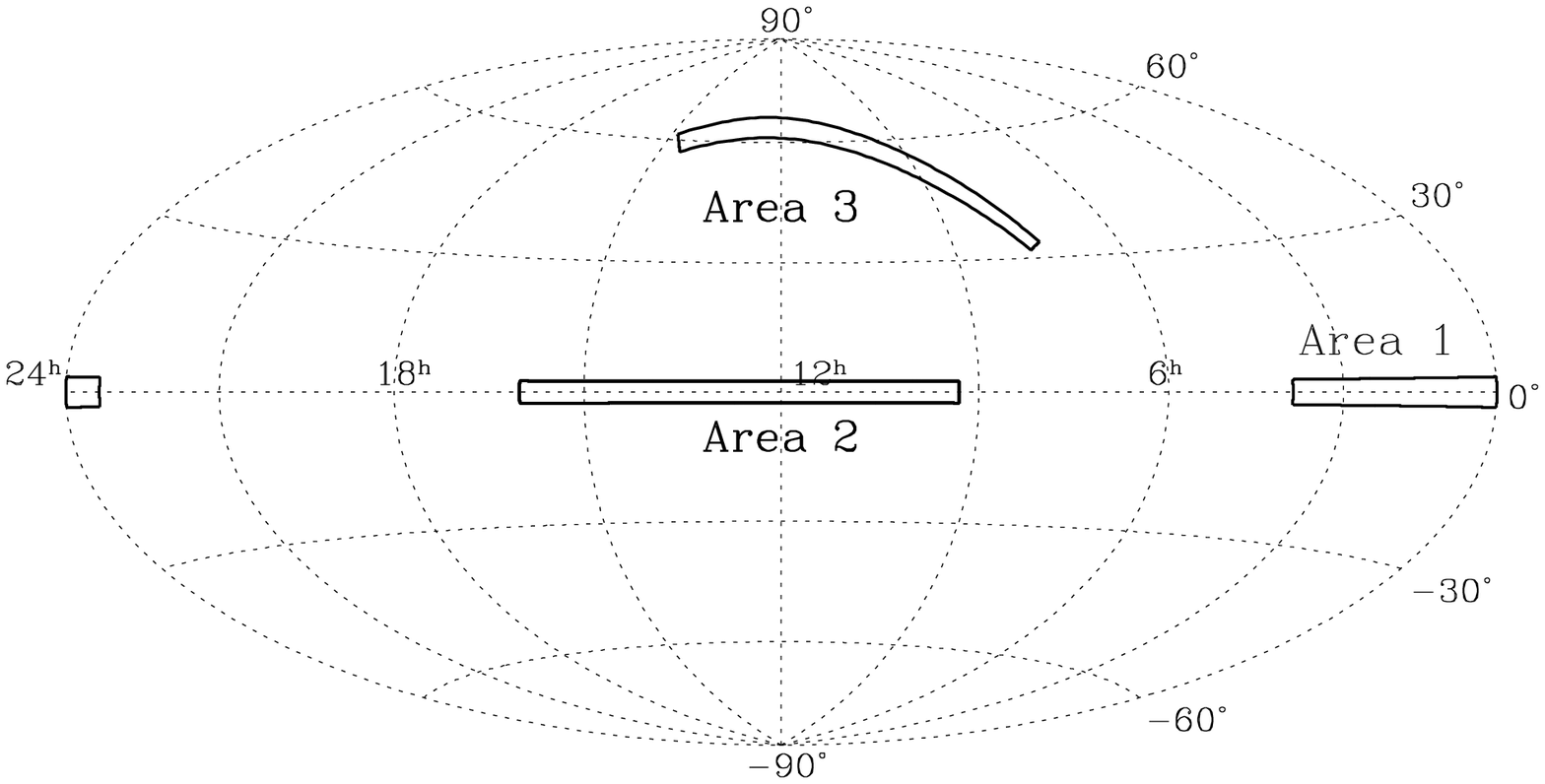}}
\end{tabular}
\caption{An Aitoff projection in celestial coordinates of the approximate areas  covered by the three SDSS regions from which we measure the two-dimensional genus. Coordinate limits for these regions are described in the text. }
\label{fig:data}
\end{centering}
\end{figure}

\begin{figure} 
\begin{centering}
\hspace{-1cm}{\epsfxsize=6truecm \epsfysize=2truecm \epsfbox[0 50 550 200]{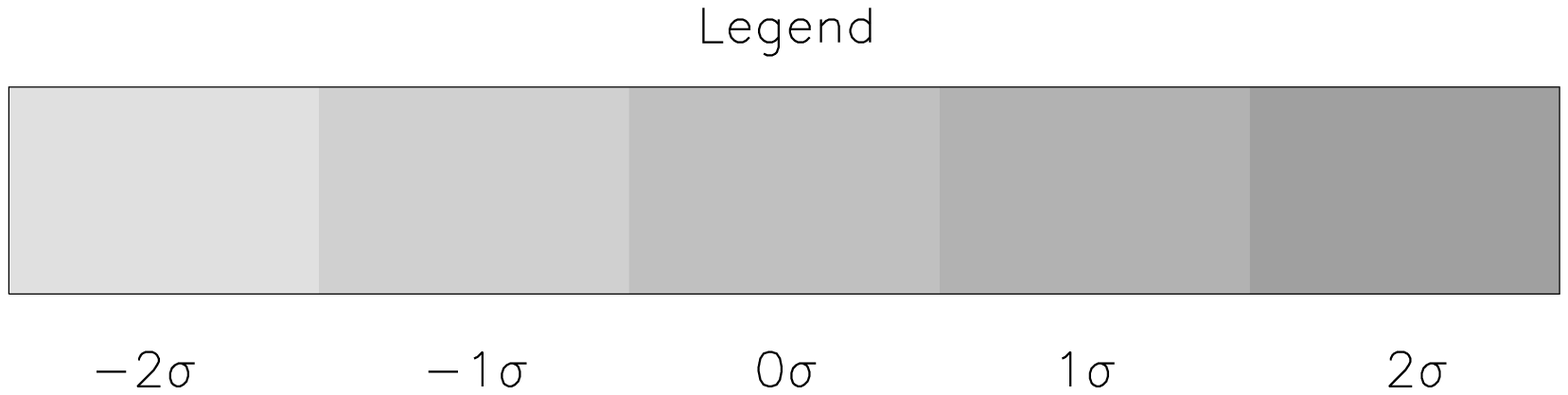}} 
\begin{tabular}{ccc}
{\epsfxsize=5truecm \epsfysize=5truecm \epsfbox[200 250 400 450]{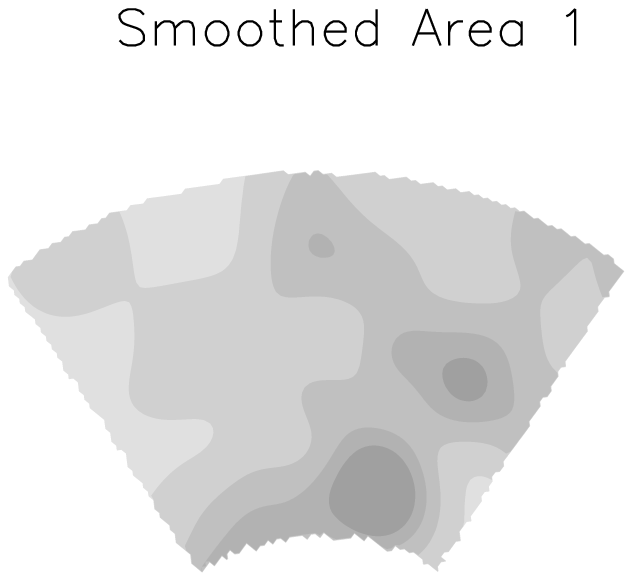}} &
{\epsfxsize=5truecm \epsfysize=5truecm \epsfbox[200 250 400 450]{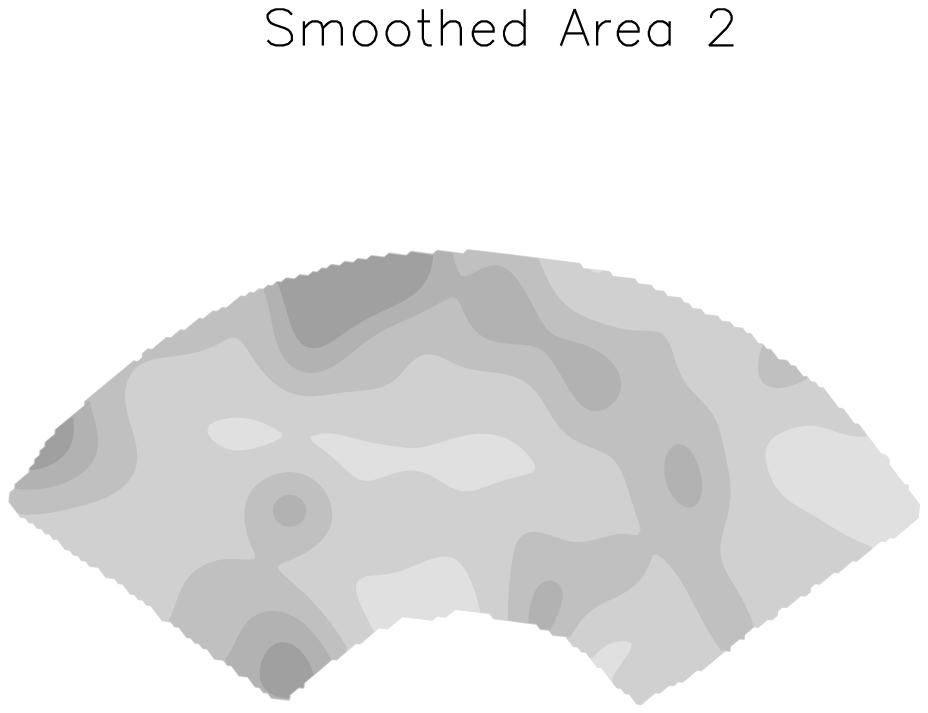}} &
{\epsfxsize=5truecm \epsfysize=5truecm \epsfbox[200 250 400 450]{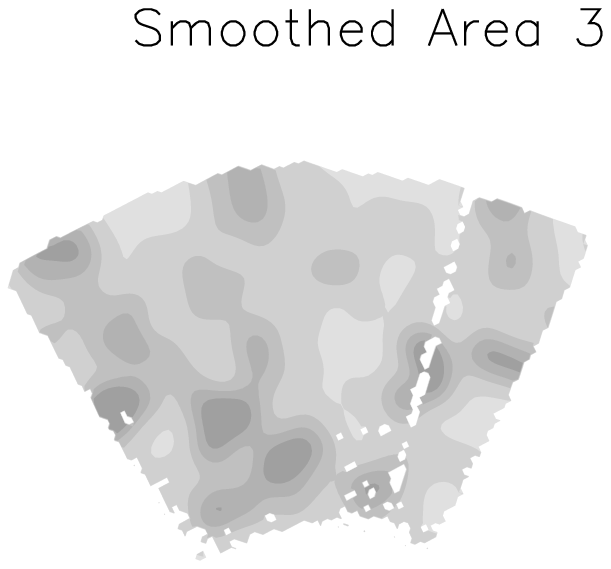}} \\
{\epsfxsize=5truecm \epsfysize=5truecm \epsfbox[150 360 470 680]{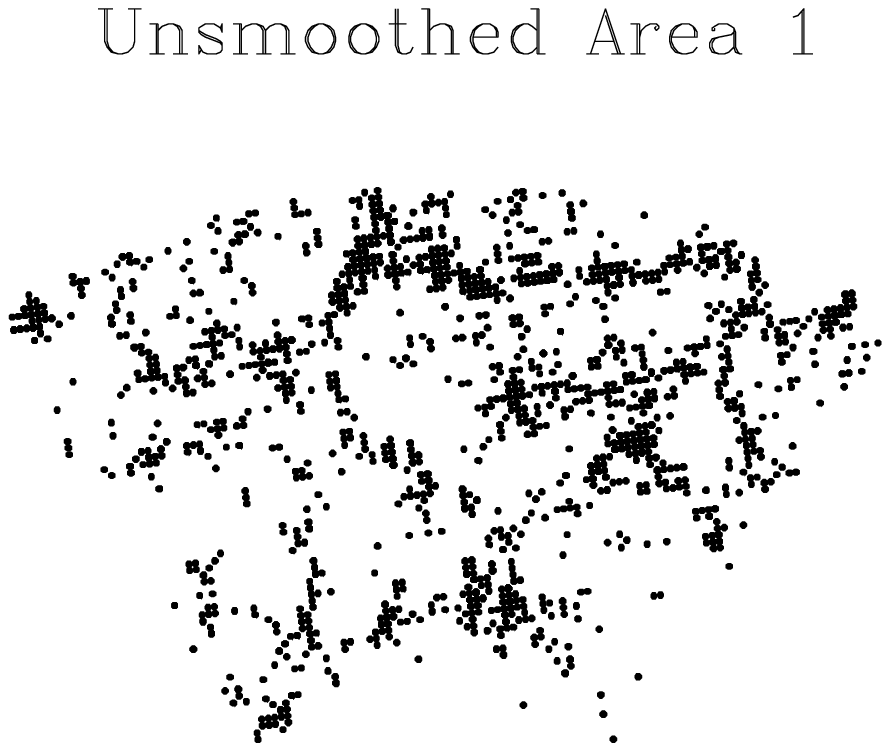}} &
{\epsfxsize=5truecm \epsfysize=5truecm \epsfbox[150 360 470 680]{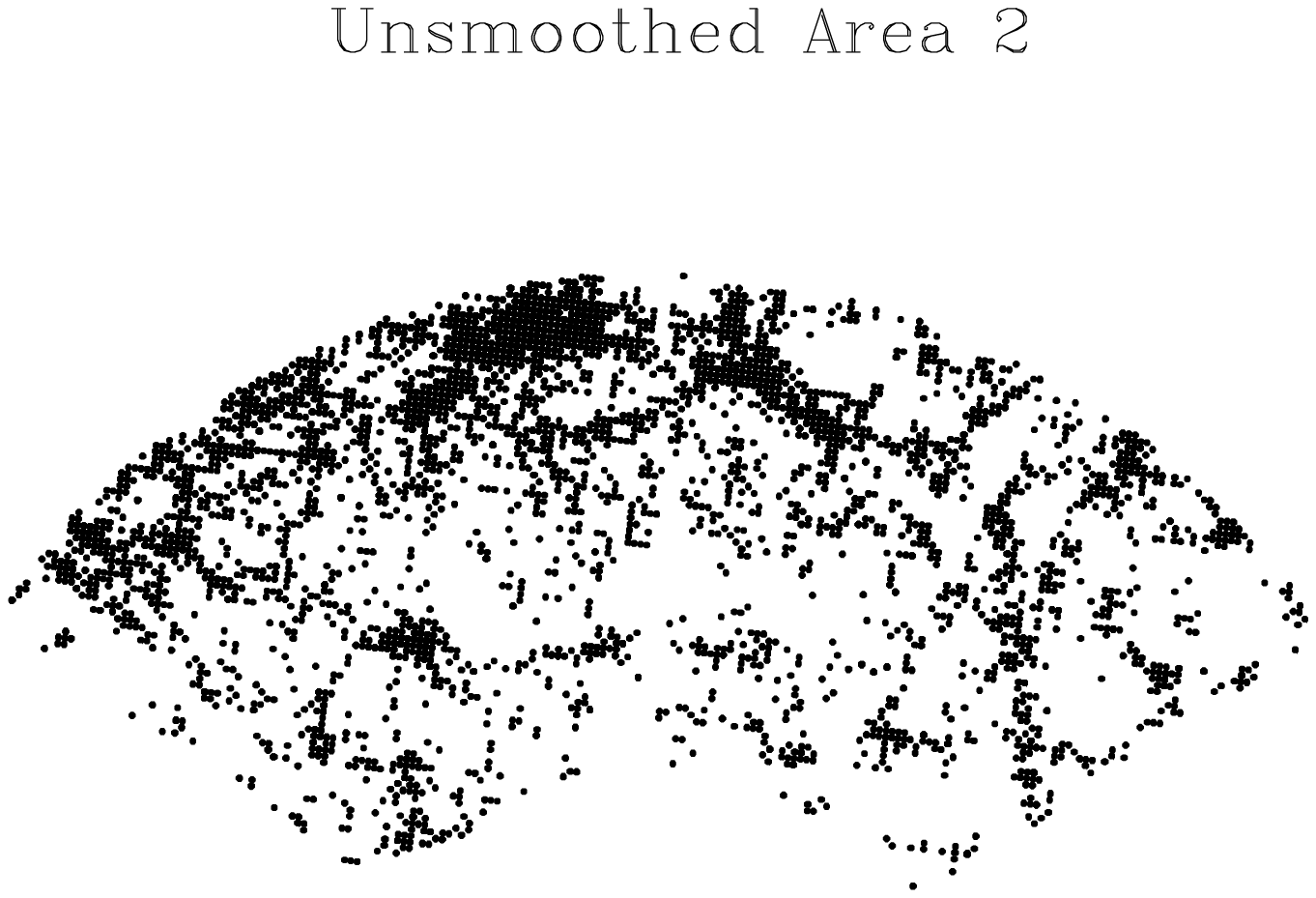}} &
{\epsfxsize=5truecm \epsfysize=5truecm \epsfbox[150 360 470 680]{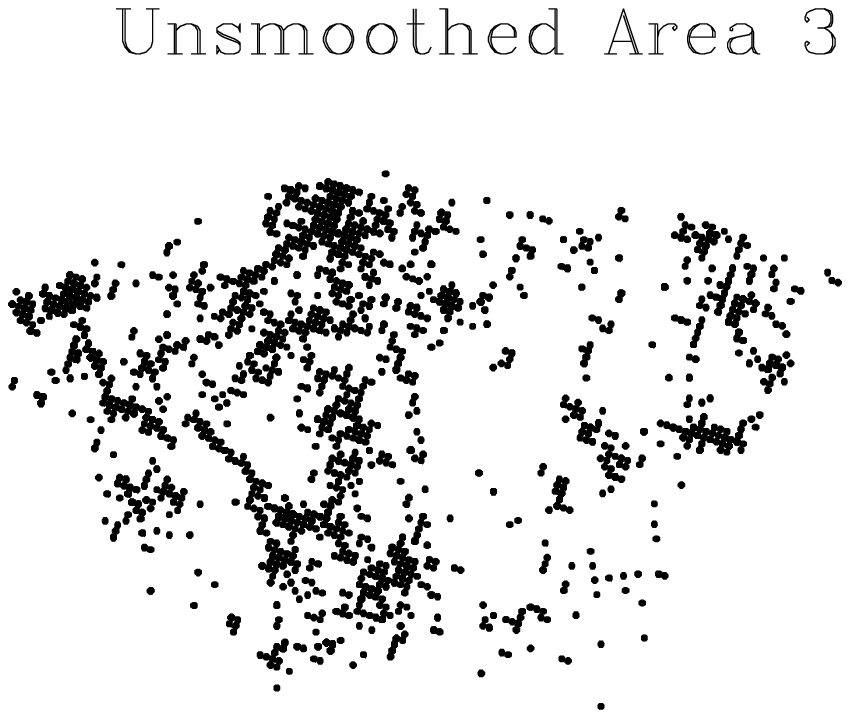}} \\
\end{tabular}
\caption{Smoothed and unsmoothed maps of the density of objects in the
SDSS slices. Darker contours correspond to denser regions.  In the
smoothed plots, the darkest contours are 2$\sigma$ over-dense regions
and the lightest contours are the 2$\sigma$ under-dense regions. The
2D smoothing kernel is a Gaussian with dispersion $\lambda
=5h^{-1}$Mpc. Small differences in the qualitative appearance of the
smoothed and unsmoothed maps are due to incompleteness of the
survey. This incompleteness and the varying thickness of the slices
with redshifts are accounted for in the smoothed maps.  In Area 3 a
region of incomplete spectroscopy causes the narrow underdensity which is
excluded from our analysis.}
\label{fig:datasmooth}
\end{centering}
\end{figure}

The SDSS is a wide-field photometric and spectroscopic survey. The
completed survey will cover approximately 10,000 square degrees. CCD
imaging of 10$^8$ galaxies in five colors and follow-up
spectroscopy of 10$^6$ galaxies with $r^{\prime}<17.5$ will be
obtained. York et al. (2000) provides an overview of the SDSS, Stoughton et
al. (2002) describes the early data release (EDR) and details about the
measurements.
Technical articles providing details of the SDSS include descriptions
of the photometric camera (Gunn 1998), photometric analysis (Lupton et
al. 2002), the photometric system (Fukugita et al. 1996; Smith et al. 2002), 
the photometric monitor (Hogg et al. 2001), astrometric
calibration (Pier et al. 2002), selection of the galaxy spectroscopic
samples (Strauss et al. 2002; Eisenstein et al. 2001), and spectroscopic
tiling (Blanton et al. 2001).

We examine a sub-sample that includes 63,125 galaxies with redshifts
in the range $0.02 < z < 0.20$ and absolute magnitude limited to
$-22.0 < M_r <-19.0$. Some of these data have already been released to
the community in the Early Data Release (see Stoughton et al. 2002).
A thorough analysis of possible systematic uncertainties in the galaxy
samples is described in Scranton et al. (2001).  As we demonstrate,
the sample is now large enough for meaningful two-dimensional genus
measurements to be made.

To a good approximation, the sample we analyze consists of three great
circle wedges, each approximately 5 degrees thick, with lengths of 70,
100, and 70 degrees.  The physical
thickness of these wedges is 8$h^{-1}$Mpc at the inner edge and
22$h^{-1}$Mpc at the outer edge. In detail, the three areas of
interest are the stripe along the Celestial Equator in the Southern
Galactic Cap (hereafter Area 1), which covers the region $ 350^{\circ}
< \alpha < 58^{\circ}$, $-2.5^{\circ} < \delta < 2.5^{\circ}$, a
region along the Celestial Equator in the Northern Galactic Cap
(hereafter Area 2), $ 140^{\circ} < \alpha < 240.5^{\circ}$,
$-2.5^{\circ} < \delta < 2.5^{\circ}$, and a further region at higher
latitudes in the Northern Galactic Cap which is better expressed in
survey coordinates, $\lambda$ (longitude) and $\eta$ (latitude) with $
-17^{\circ} < \lambda < 53^{\circ}$, $80.75^{\circ} < \eta <
86.25^{\circ}$ (hereafter Area 3). The survey coordinates are great
circles on the sky and are explained in detail in Stoughton et
al. (2002). The three different areas are shown in Figure
\ref{fig:data} and the area and number of galaxies in each area is
given in table \ref{tab:rank}.

As the survey is not finished, the angular selection function is
complicated. An algorithm to quantify the completeness, i.e. the
fraction of galaxies that have been observed in any given
region, has been developed and is described in Tegmark, Hamilton, \& Xu
(2001). The completeness for any given ($\alpha, \delta$) coordinate
is returned. The completeness within the regions that have been
observed is typically $> 90\%$.

We construct a volume-limited sample from the SDSS. This choice
reduces the number of galaxies available, but it has several important
advantages: the radial selection function is approximately uniform,
thus the only variation in the space density of galaxies with radial
distance is due to clustering. Therefore, no weighting scheme is
required when constructing the density field; thus, the analysis is
more straight forward. In addition to having nearly uniform selection
function, volume-limited samples have a constant mix of galaxy types
with redshift. The importance of this uniformity is demonstrated in
this paper; we find that the density fields traced by the reddest and
bluest galaxies do indeed exhibit different topological
characteristics.

\begin{figure} 
\begin{centering}
\begin{tabular}{c}
{\epsfxsize=8truecm \epsfysize=8truecm \epsfbox[40 120 600 700]{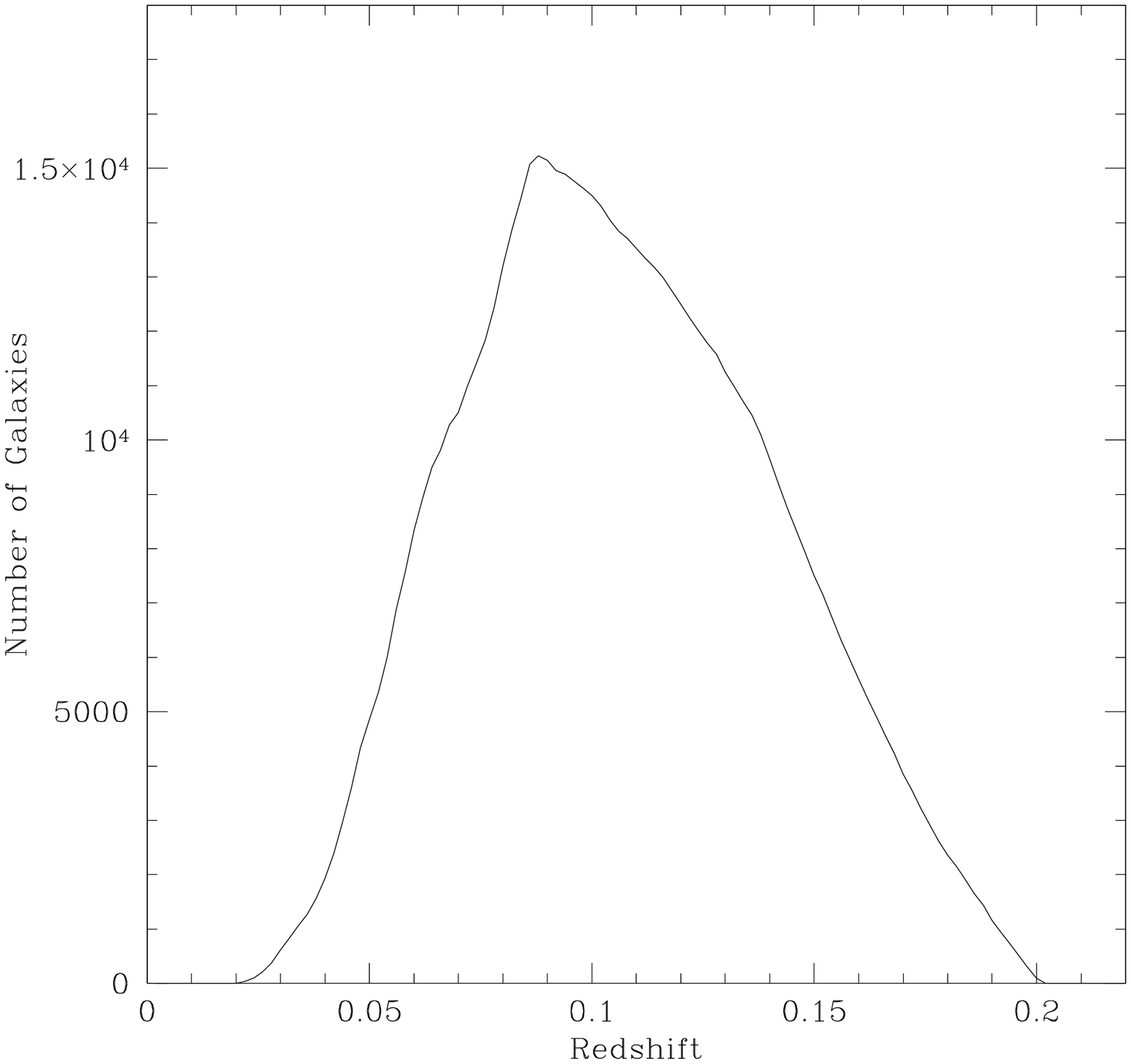}}
\end{tabular}
\caption{Number of galaxies in a volume-limited sample, selected from
all the SDSS galaxies, as a function of the redshift limit. The sample size
peaks at redshift $z=0.088$, which corresponds to an absolute
magnitude limit of $M_{\rm lim}=-19.86$ in the $r^*$ band. }
\label{fig:vol}
\end{centering}
\end{figure}

We form a volume-limited sample that extends to $z_{\rm
max}$=0.088. This limit includes the maximum number of galaxies in the
sample. In Figure \ref{fig:vol} we plot the number of galaxies in a
volume-limited sample as a function of redshift. The k-corrections we
use are described in Blanton et al.  (2002).  The redshift limit of
$z_{\rm max}$=0.088 corresponds to a limiting absolute magnitude of
$M_{\rm lim}=-19.86 + 5\log h$ (hereafter we drop the $5\log h$ term
when discussing absolute magnitudes) in the $r^*$ band, assuming an
average value for the k-correction.  Following our previous paper
(HVG02) and to maintain consistency with the Hubble Volume simulation,
which we describe below in Section \ref{sec:sims}, we adopt a
$\Omega_{\rm m}=0.3, \Omega_{\Lambda}=0.7$ cosmology when converting
redshift into comoving distances. For this cosmology, the redshift
limit of $z_{\rm max}$=0.088 corresponds to a comoving distance of
260$h^{-1}$Mpc.  There are 15,230 galaxies in the volume-limited
sample to this distance.  However, some of these survey data do not
lie in any of the three areas that we consider and we only include
galaxies at distances larger than 100$h^{-1}$Mpc because the
wedge-shaped three-dimensional survey volume is very thin out to this
distance. The final volume-limited sample contains 11,884 galaxies.

\subsection{The Simulation}
\label{sec:sims}

To test the robustness of our statistical methods, to estimate
uncertainties in our measurement, and to test the currently best
fitting variant of CDM, we use mock surveys drawn from the Virgo
Consortium's Hubble Volume $z=0$ $\Lambda$CDM simulation. For more
details on Virgo Consortium Simulations see Frenk et al. (2000) and 
Evrard et al. (2002). This
particular simulation contains 1 billion mass particles in a cube with
side 3,000$h^{-1}$Mpc. The cosmological parameters of the simulation
are $\Omega_{\rm b}=0.04, \Omega_{\rm CDM}=0.26, \Omega_{\Lambda}=0.7,
h=0.7, \Gamma=0.17$, close to the values suggested by median
statistics analysis (Gott et al. 2001). This simulation does not
include any hydrodynamics or other detailed physics but we smooth on
5$h^{-1}$Mpc scales. On this and larger scales, the dark
matter fluctuations might be expected to trace the clustering pattern
of galaxies.  The clustering amplitude of the dark matter particles
closely matches that of present day, optically selected galaxies so no
biasing has been applied.  In any case, estimation of genus as a
function of area fraction reduces the sensitivity of the topology to
galaxy bias, since it requires only a monotonic relation of the
smoothed mass and galaxy density fields to get the same genus curve,
not equality of the fields or a linear relation between them.

We construct samples that have the same geometry as the current SDSS
data.  We sparse sample the dark matter particles to match the
number density of galaxies in the SDSS.

\section{The Genus Statistic}
\label{sec:genus}

The genus statistic is a quantitative measure of the topology of
regions bounded by isodensity surfaces. In two-dimensions, those
surfaces are simply the set of curves that separate high and low density
areas.  Following Melott et al. (1989), we define the 2-D genus to be

\begin{equation}
G_2 = {\rm number \; of \; isolated \; high \; density \; regions \; - \; number \; of \; isolated \; low \; density \; regions.}
\end{equation}

The genus of a contour in a two-dimensional density distribution can also be calculated using the Gauss-Bonnet theorem, in the two dimensional form, using
\begin{equation}
G_2 = \frac{1}{2 \pi} \int C dS
\end{equation}
where the line integral follows the contour and $C = r^{-1}$ is the
inverse curvature of the line enclosing a high or low density
region and $G_2$ is proportional to the Euler-Poincar\'{e} characteristic of
the curve. This curvature can be positive or negative depending on
whether a high or low density region is enclosed. The genus of a
contour enclosing an isolated high density region is positive, whereas 
the genus of a contour enclosing a low density region is
negative. This expression allows contributions from contours around
regions that do not lie fully within the boundaries of the survey.

For a Gaussian random field, the genus has a simple analytic form. The
genus per unit area is expressed as
\begin{equation}
g_2 = \frac{G_2}{\rm Area} = A \nu e^{- \nu^2/2},
\label{eq:genus}
\end{equation}
where $\nu$ is the threshold value, above which a fraction, $f$, of the area has a higher density
\begin{equation}
f = (2\pi)^{-1/2} \int^{\infty}_{\nu} e^{-x^2/2} dx
\end{equation}
For example, at $\nu=1$ we evaluate the topology of a contour that
encloses 16\% of the area with highest density and at $\nu =0$, the contour
is drawn at the median density.
The value of the amplitude, $A$, depends on the shape of the smoothed power
spectrum,
\begin{equation}
A = \frac{<k^2>}{2(2 \pi)^{3/2}}
\end{equation}
and $<k^2>$ is the square of the wavenumber, $k$, averaged over the
two-dimensional power spectrum.

The steps to calculating the genus are as follows:
\begin{itemize}
\item Construct a random catalog of points for each SDSS area that
has the same angular and radial selection function as the galaxies
but with much higher number density.
\item Bin the galaxies (each area is considered
separately) and the random points on a two-dimensional grid. We
have 256$^2$ cells in each grid. Calculate the density of data and
random points in each cell.
\item Smooth the data and random density field with a Gaussian kernel
\item Divide the smoothed data density field by the smoothed random
density field.
\item Mark the cells that lie outside the survey (i.e., the cells with
density=0 in the unsmoothed random density field) with a negative
value. These cells are then ignored by CONTOUR2D.
\item Run CONTOUR2D to compute the genus at several density thresholds.
\end{itemize}

CONTOUR2D (Melott 1989) \footnote{Note that the version of CONTOUR2D
published in Melott (1989) contains a sign error in line 19 of the OUT
subroutine.}  uses a two-dimensional variant of the angle deficit
algorithm described by Gott et al. (1986). The density contour is
specified by the fractional area contained in the high density region
of the smoothed density field. The program classifies all cells as
high or low density then evaluates the genus of the contour surface by
summing the angle deficits at all vertices that lie on the boundary
between the high and low density regions.

\subsection{The Smoothing Length}

We smooth the data and random point density distributions by
convolving the density field with a Gaussian of the form
$\exp(-r^2/2\lambda^2)$, where $\lambda$ is the smoothing scale.
Note that this definition of the Gaussian smoothing length $\lambda$
differs from some earlier papers on the genus statistic (some earlier
papers use $\exp[-r^2/\lambda^2]$).

Vogeley et al. (1994) describe a series of desiderata that the
smoothing length should satisfy.  First, the smoothing length should
be larger than the grid cell.  For smoothing, we bin the galaxies in
each of the three SDSS regions onto a $256^2$ grid with side length
$575h^{-1}$Mpc, thus the grid cell size is of order $2h^{-1}$Mpc.

The second criterion is that we require the variance due to
clustering, $\sigma_{\lambda}$, on the smoothing scale, $\lambda$, be
larger than the Poisson error on this same scale. This ensures that we
are dominated by clustering signal rather than shot noise. For top-hat
smoothing with radius $\lambda$ (the results for a Gaussian smoothing
are of the same order), the constraint would be
\begin{equation}
\frac{\delta N_{\rm clustering}}{N} > \frac{\delta N_{\rm poisson}}{N}
\end{equation}
\begin{equation}
\sigma_{\lambda} > \frac{1}{\sqrt{N}} = \frac{1}{\lambda \sqrt{\pi \bar{\mu}}}
\end{equation}
\begin{equation}
\lambda > ( \sigma_{\lambda} \sqrt{\pi \bar{\mu}} )^{-1}
\end{equation}
where $\bar{\mu}$= $N$/A, the number of galaxies in the sample divided by the 
area of the sample. For a smoothing scale of 5$h^{-1}$Mpc this value is small,
of order 0.5$h^{-1}$Mpc.

\begin{table}
\begin{centering}
\begin{tabular}{cccccccccc}
Data & Area & $N_{\rm gal}$ & $N_{\rm res}$ & Amp & Rank & Significance & Shift & Rank & Significance \\  \hline
Area 1 & 35,000 & 2,343 & 435 & 8.6 & 13 & 0.62 $< 1\sigma$ & 0.05 & 14 & 0.66 $< 1\sigma$ \\
Area 2 & 54,000  & 6,970 & 685 & 14.6 & 4 & 0.19 $\sim 1\sigma$ & -0.15 & 3 & 0.14 $\sim 1\sigma$ \\
Area 3 & 35,000 & 2,571 & 435 & 7.4 & 3 & 0.14 $\sim 1\sigma$ & 0.2 & 17 & 0.81 $\sim 1\sigma$ \\ \hline
Combined & 124,000  & 11,884 & 1555 & 28.1 & 4 & 0.19 $\sim 1\sigma$ & -0.05 & 8 & 0.38 $< 1\sigma$ \\ \hline
\end{tabular}
\caption{Properties and rank of the amplitude and shift for the genus
curves from the three areas and the summed genus curve as compared to
the 20 realizations from the simulation. The area is given in units of
$h^{-2}$Mpc$^2$ and the number of resolution elements is calculated
assuming a smoothing length 5$h^{-1}$Mpc. The samples are
volume-limited with $z_{\rm max}=0.088$ corresponding to a magnitude
limit of M$_{\rm lim}=-19.86$ in the r* band. The rank is that of each
dataset with respect to the 20 mock catalogs, i.e. a ranking of 9
means there are 8 mock catalogs with smaller values and 12 with larger
values of the feature we are testing.}
\label{tab:rank}
\end{centering}
\end{table}

The third criteria is that we want the smoothing length
to be larger than the average inter-galaxy separation. 
This avoids, for example, erroneously treating a single galaxy as a ``cluster.''
This is satisfied if
\begin{equation}
\lambda > ( \bar{\mu})^{-1/2}.
\end{equation}
For the three SDSS regions, the value of $\bar{\mu} \sim 0.1 $,
therefore $\lambda$ is required to be larger than 3$h^{-1}$Mpc. At the
inner edge of the Survey, the wedge shape is thinner and hence the
value of $\mu$ is less than the average value. At the inner edge of
the wedge, the value of $\mu^{-1/2} \sim 4.5 h^{-1}$Mpc.

The number of resolution elements in the survey depends on the
smoothing length applied. We wish to make this number as large as
possible.  The number of resolution elements, $N_{\rm res}$, in
two-dimensions is roughly
\begin{equation}
N_{\rm res} = \frac {A_{\rm sample}}{\pi \lambda^2}
\end{equation}
therefore it is advantageous to use the smallest value of $\lambda$
consistent with the shot noise constraints possible. We choose a
smoothing length of 5$h^{-1}$Mpc for most of our analysis.  Table
\ref{tab:rank} lists the number of resolution elements for each
area. Using a 5$h^{-1}$Mpc smoothing length we have at least 400
resolution elements in the measurement of the genus from each area.
For comparison, the 2dFGRS extends to greater depth so even with a
smoothing length of 10$h^{-1}$Mpc, HVG02 used of order 400 resolution
elements in each of the NGP and SGP slices.

\subsection{Errors}

We use the simulations as a method for estimating the errors on the
genus curve measured from the SDSS. The two-point clustering amplitude
of the $\Lambda$CDM Hubble Volume simulation is close to that of
present day, optically selected galaxies (see Hoyle et al. 1999).

In HVG02, we investigated calculating the full covariance matrix and
employing that matrix when $\chi^2$ fitting the data to a Gaussian random
curve. We found that 20 mock catalogs were not sufficient to
adequately describe the full covariance matrix, even if the higher
order clustering was truly described by the $\Lambda$CDM
simulation. Therefore we use the 20 mock
catalogs only to estimate the variance of each genus measurement.
As the amplitude of the SDSS genus curve closely matches
that of the simulation, it seems reasonable to use the errors from the
simulation as estimates for the errors in the data.

In the figures below we plot estimated uncertainties for every other
point in the genus curve. This choice reflects the strong covariance
of genus estimates and makes the plots easier to read.

\subsection{Combining the Genus Curves}
\label{sec:comb}

To combine estimates of the genus from different samples, whether from
sub-regions of the same survey, or from different surveys, one should
not simply add the genus curves $G(\nu)$. A better procedure is to
compute $G(\rho)$ for each sample, then combine the results into a
summary statistic $G(\nu)$, using the relationship $\nu(\rho)$ derived
from the combined sample. This procedure, detailed in this section
below, avoids errors caused by fluctuations in the mean
density and the density probability distribution.

CONTOUR2D computes the genus at a set of specified area fractions
for the low-density region enclosed by the isodensity contour.
This fraction is parametrized by $\nu$, as in eq. (4).  The code
iterates to find the density threshold corresponding to this area
fraction, draws the contour, computes the genus of this contour and
returns $\rho$ and $G$ for each area fraction. The value of the
density threshold $\rho(\nu)$ corresponding to fixed $\nu$ can vary
among different samples under consideration due to variations in
large-scale structures present in the samples.  Thus, rather than
naively summing genus curves for different samples at each
value of $\nu$, we sum the values of the genus at matching density
thresholds $\rho$ when calculating the total genus curve of the survey.

The procedure is as follows: First the area fraction and the genus
is found for each area as a function of $\rho$.  We compute the genus
for a large number of threshold values, so that the curves may be
interpolated to later find $G(\rho)$ at any value of $\rho$.  We then
calculate the average area fraction at each $\rho$ value by summing
the three individual area fractions, weighted by the areas of each
region. The total genus at each density value is found by summing the
three genus values as the total genus takes into account the different
areas. This yields the total genus of the SDSS as a function of the
area fraction. The final step is to convert the area fraction back
into $\nu$ values. The resulting curve is identical to the curve that
would be obtained by binning the data for all of the sub-regions onto
a larger grid and simultaneously computing the genus for all three
regions.

Note carefully that, at the negative $\nu$ end of the genus curve,
changes in the area fraction lead to very small changes in the
density threshold (i.e., the gradient of the density field is quite
shallow).  As a result, in some cases the genus curves of the combined
sample are neither computed nor plotted at very low values of $\nu$.
In contrast, at the high $\nu$ end, changes in the area fraction can
correspond to quite large changes in the density threshold (due to the
steeper density gradient), allowing the high $\nu$ end of the genus
curve to be better sampled if we binned linearly in density.

\section{Results}
\label{sec:res}

\subsection{Genus of Volume Limited Samples of the SDSS}

\begin{figure} 
\begin{centering}
\begin{tabular}{ccc}
{\epsfxsize=5truecm \epsfysize=5truecm \epsfbox[40 120 600 700]{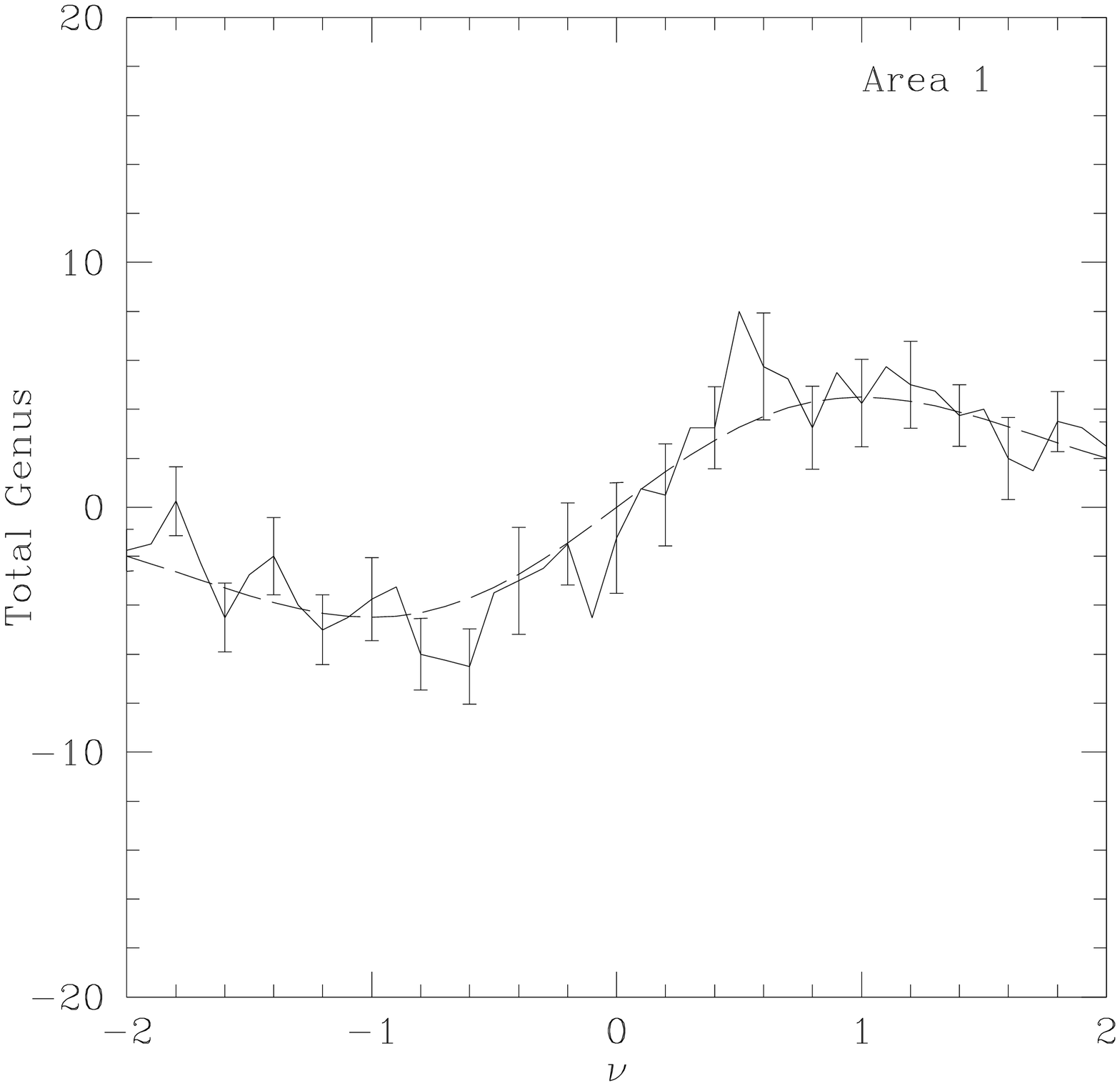}} &
{\epsfxsize=5truecm \epsfysize=5truecm \epsfbox[40 120 600 700]{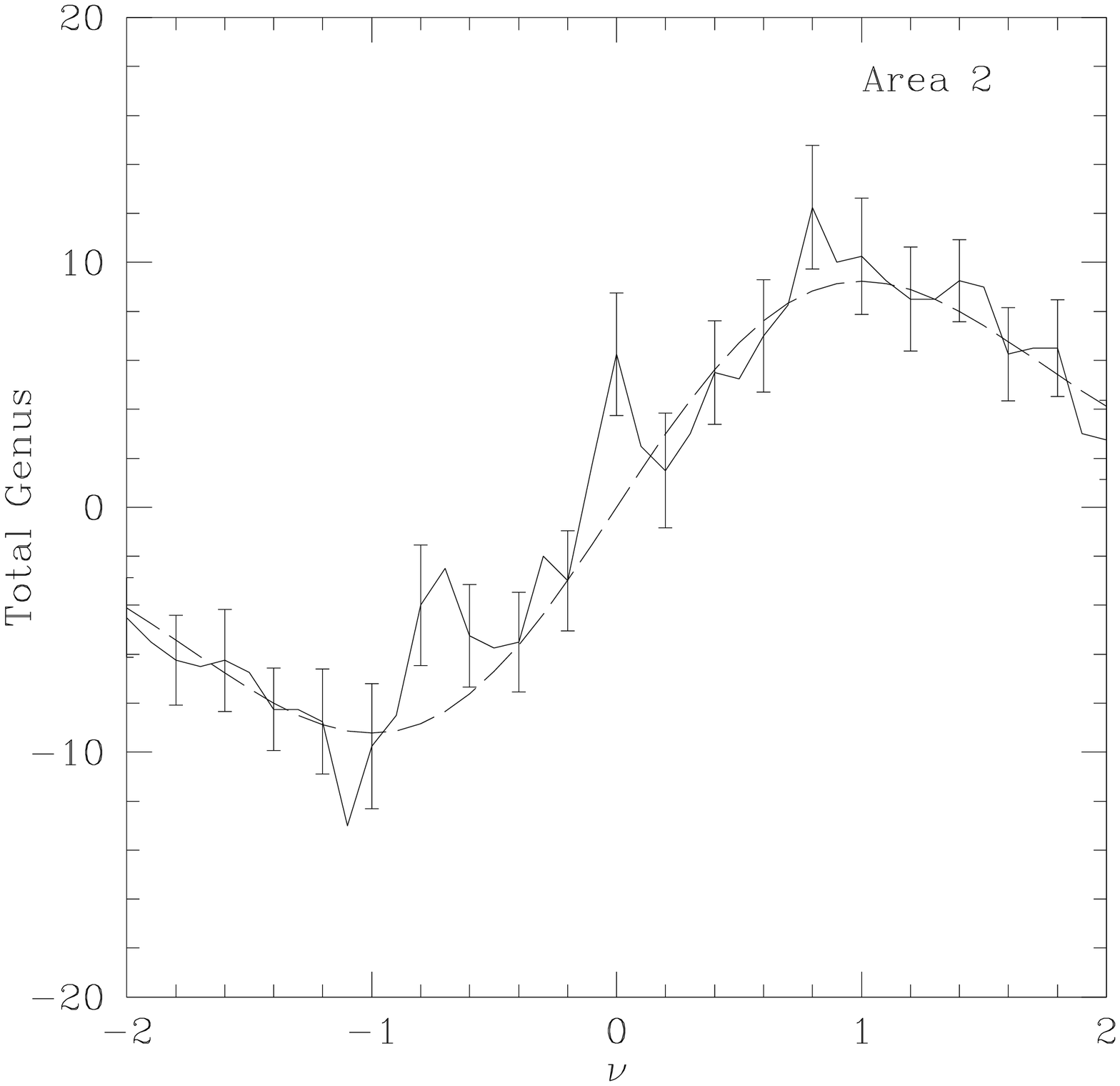}} &
{\epsfxsize=5truecm \epsfysize=5truecm \epsfbox[40 120 600 700]{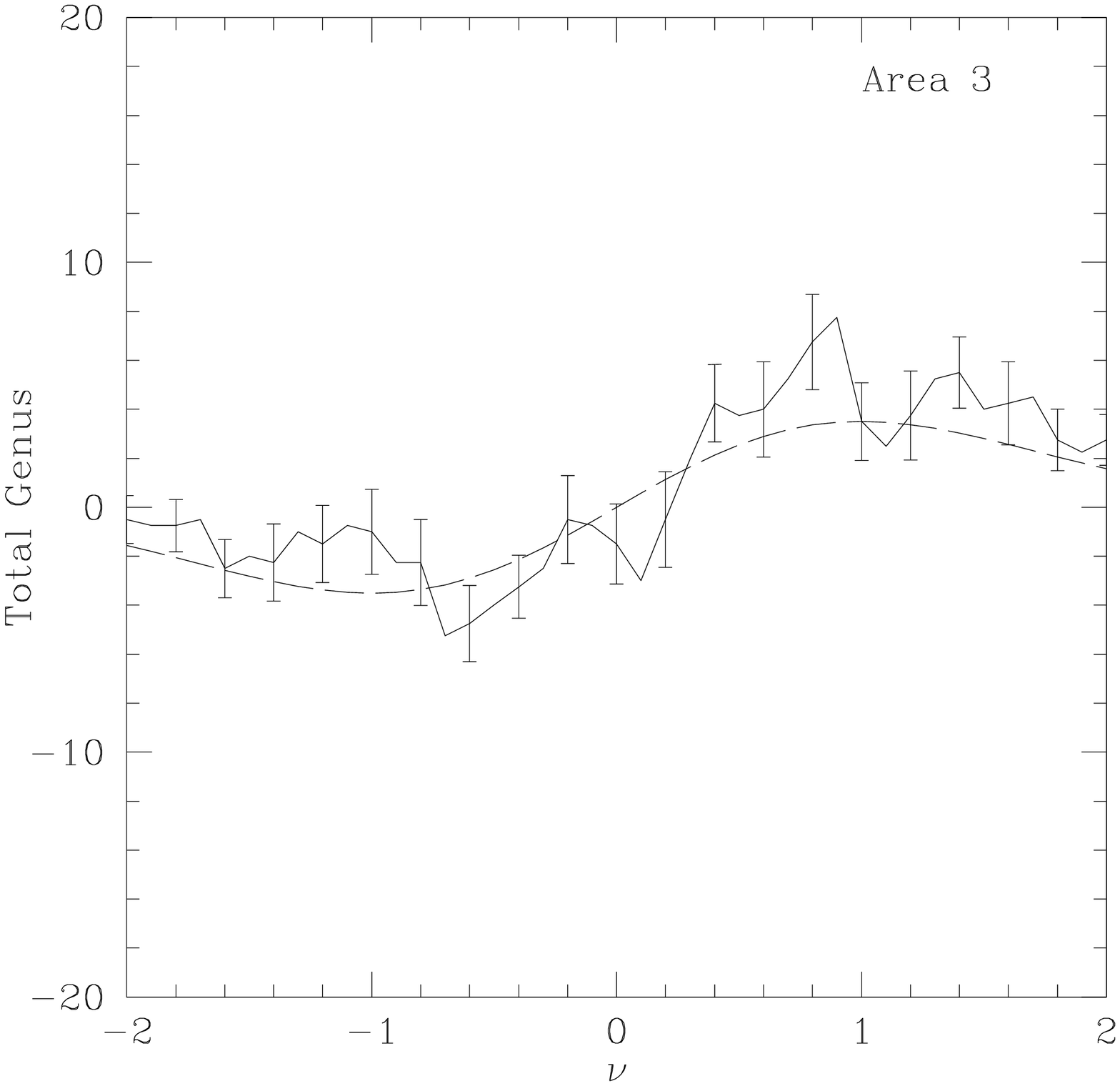}} \\
\end{tabular}
\caption{Genus curves of the three SDSS slices. A 5$h^{-1}$Mpc smoothing length is used. Errors are estimated from 20 mock catalogs drawn from the Hubble Volume $\Lambda$CDM simulation. Dashed curves show the best fit (using $\chi^2$) Gaussian random field genus (equation \ref{eq:genus}).}
\label{fig:areas}
\end{centering}
\end{figure}

\begin{figure} 
\begin{centering}
\begin{tabular}{cc}
{\epsfxsize=7truecm \epsfysize=7truecm \epsfbox[40 120 600 700]{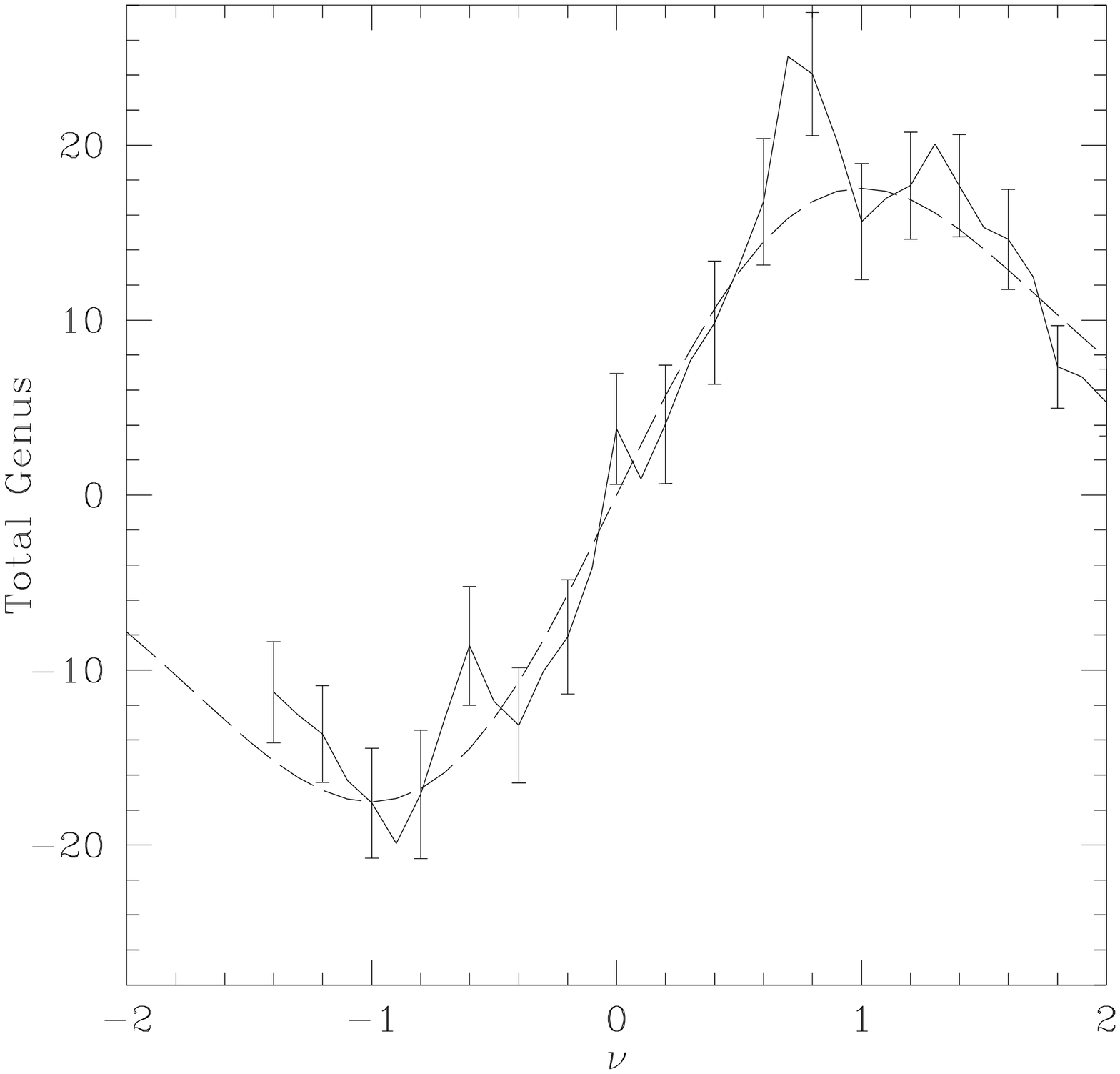}} &
{\epsfxsize=7truecm \epsfysize=7truecm \epsfbox[40 120 600 700]{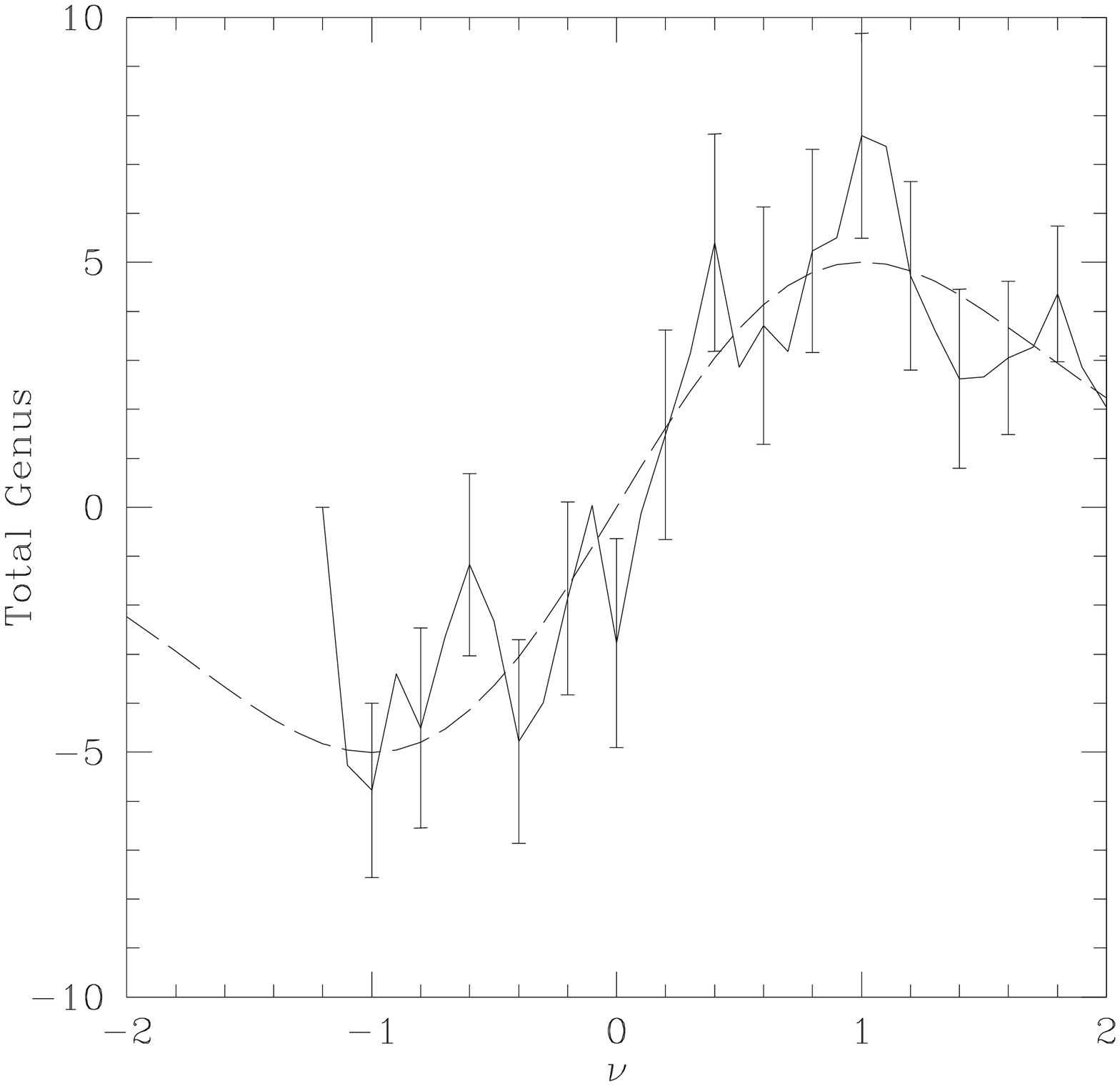}} \\
\end{tabular}
\caption{The total genus curve (combined as described in Section \ref{sec:comb}) of the SDSS calculated using a 5$h^{-1}$Mpc smoothing length (left hand plot) and a 10$h^{-1}$Mpc smoothing length (right hand plot). Dashed curves show the best fit (using $\chi^2$) Gaussian random field genus (equation \ref{eq:genus}). The combined genus is not computed for the lowest values of $\nu$, as described in Section \ref{sec:comb}.}
\label{fig:smooth}
\end{centering}
\end{figure}

In Figure \ref{fig:areas} we show the genus curves from the three SDSS
areas. The data have been smoothed with a Gaussian of width 5$h^{-1}$
Mpc in all cases and we show the total genus of each area. In Figure
\ref{fig:smooth} we combine the genus curves of the three areas, as
described in Section \ref{sec:comb}, to present the total genus for
the sample of SDSS galaxies presented here.  We show the results for
two values of the smoothing length, 5$h^{-1}$ Mpc (left) and
10$h^{-1}$ Mpc (right). For comparison, we plot the best fit genus
curve of a Gaussian random field distribution, shown by the dashed
line. The genus curve found using a 10$h^{-1}$ Mpc smoothing length is
noisier due to the smaller number of resolution elements. We focus
on the 5$h^{-1}$ Mpc results for the remainder of the analysis.

As described in HVG02, we use the Mann-Whitney Rank Sum test to
compare the data to the mock catalogs. Table \ref{tab:rank} shows the
comparison with 20 mock catalogs. We measure the amplitude of the
genus curve of the data and mock catalogs by $\chi^2$ fitting to a
Gaussian curve (equation 3) with varying amplitude. The shift of the
curve is found simply by measuring the zero-crossing of the genus
curve.

\begin{figure} 
\begin{centering}
\begin{tabular}{cc}
{\epsfxsize=7truecm \epsfysize=7truecm \epsfbox[40 120 600 700]{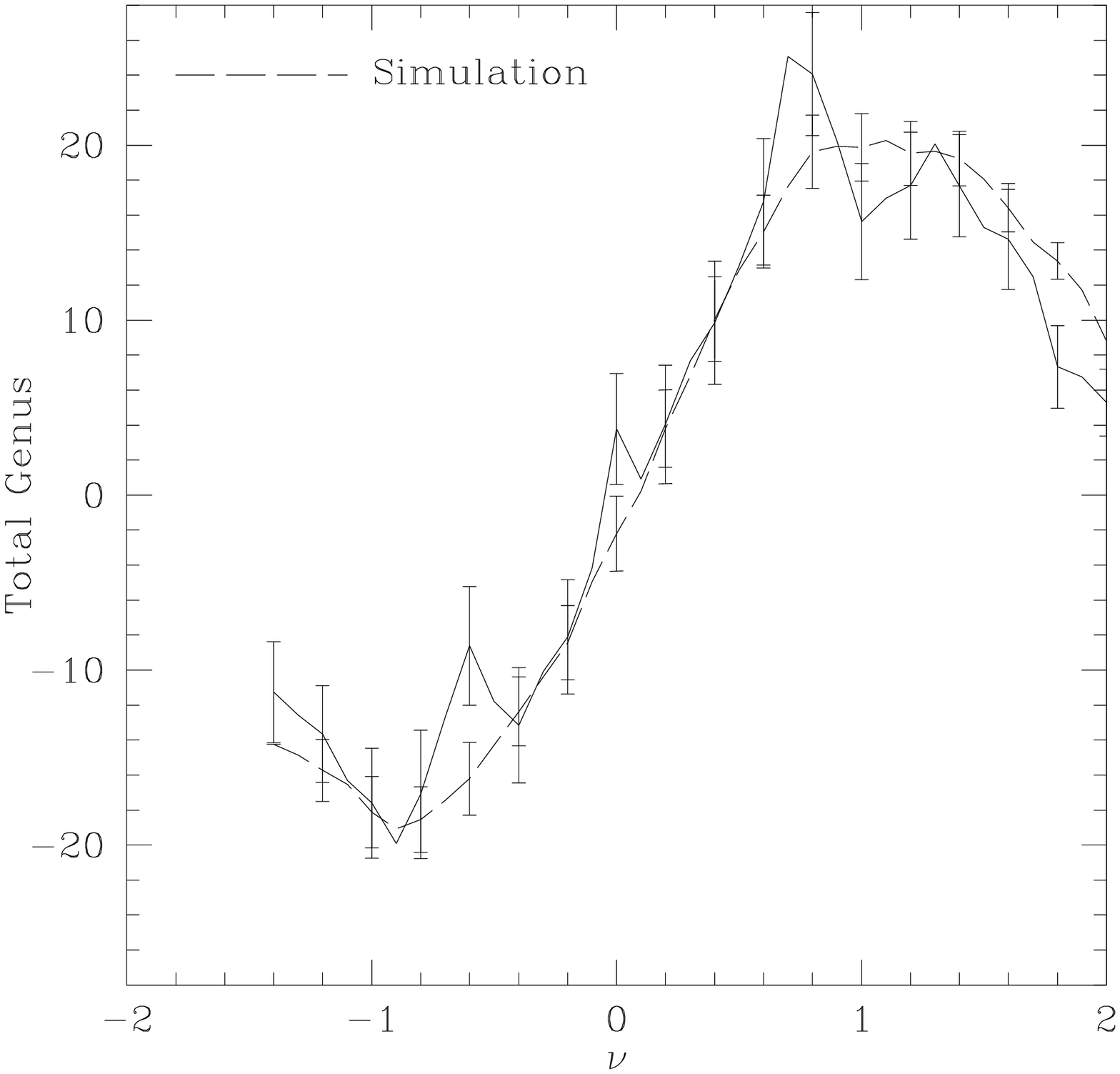}} &
{\epsfxsize=7truecm \epsfysize=7truecm \epsfbox[40 120 600 700]{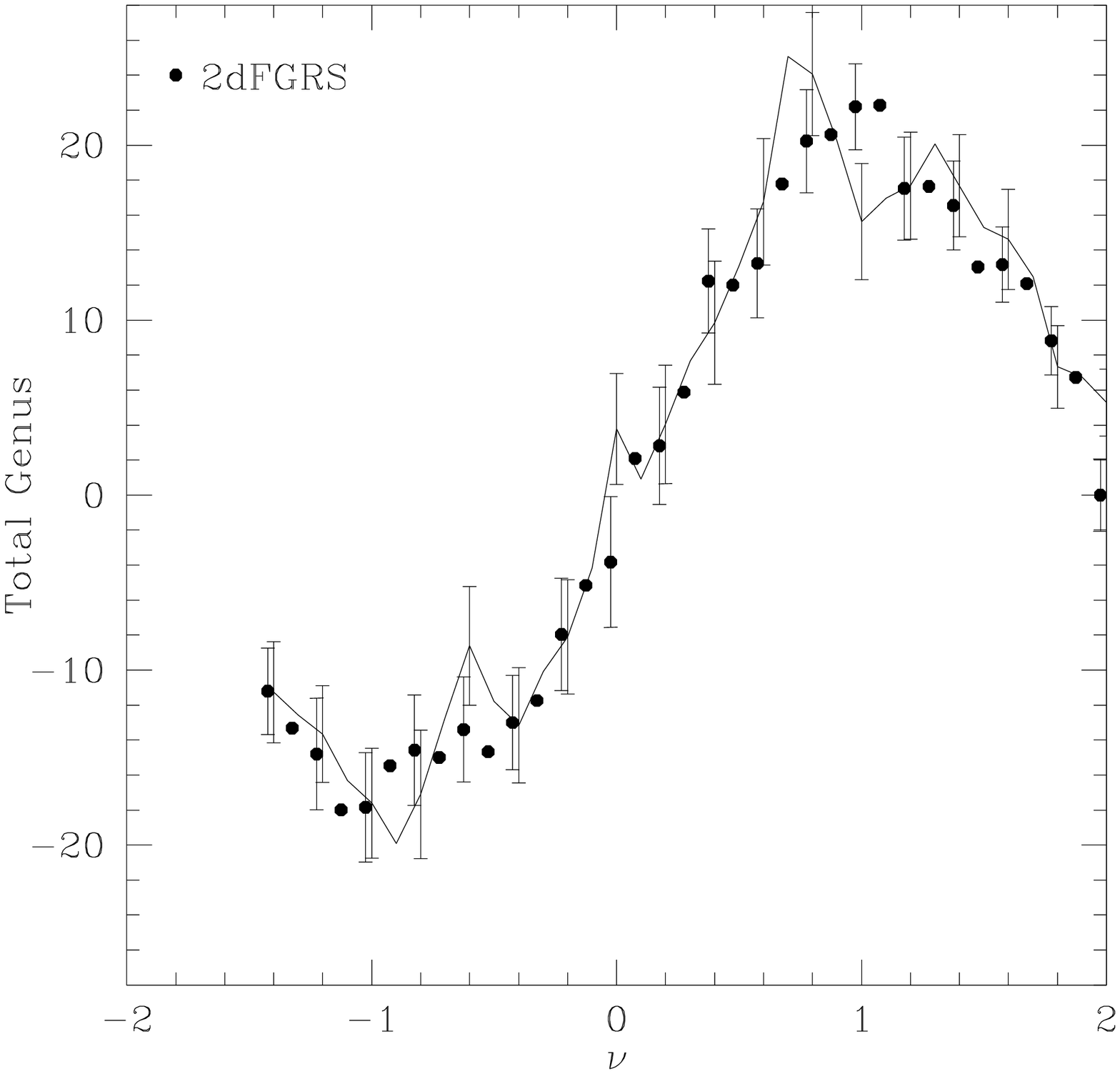}} \\
\end{tabular}
\caption{The total genus curve of the SDSS compared to the average
genus curve of the 20 mock catalogs, drawn from the Hubble Volume
$\Lambda$CDM simulation, described in Section 2.2 (left plot), and the
2dFGRS curve similar to that given in HVG02 but recombined as
described in this paper (right hand plot). Note that the 2dFGRS points
are slightly offset to the left for clarity, and that every other
2dFGRS point has an error bar attached. The points without error bars
have similar errors. A 5$h^{-1}$Mpc smoothing length is used.}
\label{fig:comps}
\end{centering}
\end{figure}

We find that the amplitude of the genus of the SDSS lies slightly
below the median genus amplitude of the simulated catalogs (rank 11).
This decrement appears to be because the genus curve of the SDSS peaks
at smaller absolute $\nu$ values than the Gaussian random curve and
the simulation, as shown in Figures \ref{fig:smooth} and
\ref{fig:comps}, respectively. For the SDSS, the width $| \nu_{\rm
max} - \nu_{\rm min}| \sim 1.6$, whereas for a Gaussian curve $|
\nu_{\rm max} - \nu_{\rm min}|=2$ and the genus measured from the
simulation, which assumes Gaussian initial conditions, also has $|
\nu_{\rm max} - \nu_{\rm min}|\sim 2$. We do not see this effect in
the 2dFGRS data. In fact, the 2dFGRS data exhibit a slight broadening
in the peak-trough separation. Neither do we see any narrowing of the
SDSS genus curve when we use a 10$h^{-1}$Mpc smoothing length,
although the smaller number of resolution elements causes the genus
curve to be noisier for $10h^{-1}$Mpc smoothing and locating the peaks
is difficult. This narrowing in the peak-trough separation in the SDSS
data using the 5$h^{-1}$Mpc smoothing length is in the opposite sense
to the results of Vogeley et al. (1994), who found that the
three-dimensional genus curve of the CfA survey was slightly broader
at the 85\% level for small choices of the smoothing length ($\lambda
< 12 h^{-1}$Mpc).

The zero-point shifts of the data are more consistent with the
simulation. Area 2 has a slight negative shift but on average the zero
crossing point agrees with the simulation. The fact that it lies at
less than zero suggests there is a slight excess of clusters (a
meatball shift) in both the data and the simulation, as seen
previously in the 2dFGRS (HVG02).

We also compare the SDSS genus curve to the 2dFGRS genus curve. The
total area of the 2dFGRS sample analyzed in HVG02 is approximately
double that of the SDSS slices analyzed here so the 2dFGRS curve has
been scaled in Figure \ref{fig:comps} to match the SDSS curve.  Here
we plot a revised genus curve for the 2dFGRS that uses the more
accurate procedure described above in section \ref{sec:comb} for
combining the genus curves of the North and South regions of that
survey. The genus curves for the full SDSS and 2dFGRS samples agree
within the estimated uncertainties.  However the cluster excess in the
2dFRGS genus curve found in HVG02 is less apparent when we compute the
genus curve for the full sample using our new more accurate method.

\subsection{Genus as a Function of Galaxy Type}

A unique feature of the SDSS project is that medium-resolution
spectroscopy and five-band CCD photometry are obtained for all the
objects in the galaxy redshift survey. Details of these
measurements may be found in Stoughton et al. (2002).

Here we examine whether the genus curve depends on the color of the
galaxy. Red, elliptical galaxies are believed to be preferentially
found in high density regions whereas blue spiral galaxies may be
found preferentially in lower density regions. This difference may lead to
quantitatively different genus curves, as discussed in Gott, Cen, \&
Ostriker (1996), where it was predicted that elliptical galaxies may show a
shift towards a meatball topology relative to spiral galaxies.
Therefore, we split the sample by $u^*-r^*$ color, which was found to
be a good indicator of galaxy type by Strateva et al. (2001). The data
divide into three equal sized samples if we split the sample at
$u^*-r^*<2.35$ and $u^*-r*>2.86$.

\begin{figure} 
\begin{centering}
\begin{tabular}{c}
{\epsfxsize=8truecm \epsfysize=8truecm \epsfbox[40 120 600 700]{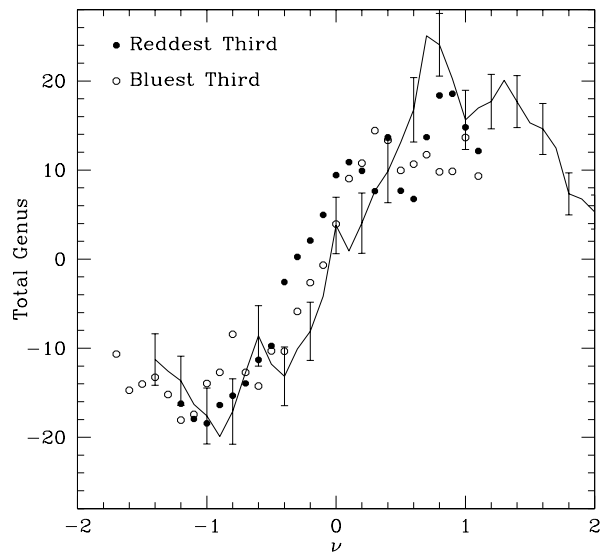}}
\end{tabular}
\caption{Genus curve of the reddest third of the
sample (filled circles) compared to the genus curve of the bluest
third of the sample (open circles) selected in $u^*-r^*$. The solid
line shows the genus curve of the full sample. For clarity, we only
plot the errors for the full sample. Errors for the red and blue
samples are similar in size. The genus curve of the reddest third
shows a meatball shift (toward $\nu<0$) relative to the
bluest third and the full sample.}
\label{fig:cols}
\end{centering}
\end{figure}

We investigate whether reducing the number of galaxies in the sample
has an effect on either the shape of the genus curve or on the size of
the errors. By sparse sampling our mock catalogs at a rate of 1 in 3,
we find that the genus curve is not changed significantly in shape,
nor do the error bars increase by more than 10\%. It appears that the
density of galaxies is high enough for the clusters and voids to be
well sampled, thus the genus curve is robust to sub-sampling of the
survey. In other words, the uncertainties on the genus curve are
caused by different structures within the samples, not by Poisson
fluctuations in the galaxy density field.

In the SDSS samples, we find (see Figure \ref{fig:cols}) noticeable
differences between the genus curves of the reddest and bluest
galaxies. The red galaxies (filled circles) appear to have a meatball
shift as compared to the bluest galaxies (open circles) and the full
sample.  In the two dimensional genus, a meatball shift occurs when
the genus remains positive -- dominated by isolated clusters -- at
negative values of $\nu$, i.e., below the median density.  This behavior for the
reddest galaxies is consistent with the prediction of Gott et
al. (1996), reflecting the higher concentration of early-type galaxies
in the highest density regions. None of the
one-in-three sampled mock catalogs from the Hubble Volume simulation
exhibit a meatball shift of this magnitude.  

While the reddest third of the galaxy sample exhibits a shift
reflective of an excess of isolated clusters, this result only became
clear after combining the three sub-regions as described in section
\ref{sec:comb}.  This effect was not clearly apparent in the genus
curves of the sub-regions.  In other words, the sample we examine is
just large enough to reveal this effect, which suggests that this
might be a statistical fluke.  We will repeat this analysis with a
larger sample as it becomes available.
The amplitude of the genus curves also appear to be suppressed when we
split the samples by color. The cause of the latter effect is not
clear.

It is important to emphasize that this measurement reveals a
difference in the topology of the distribution of galaxies segregated
by color, rather than simply reflecting the well-known
morphology-density relation (Dressler 1980; Postman \& Geller 1984).
By comparing genus curves as a function of area fraction (see
eq. 4), differences in the density probability distribution have
already been taken into account.

This observed dependence on galaxy color of the genus of the smoothed 
galaxy density field suggests that there is not a simple one-to-one 
mapping between the mass and galaxy density.
Note that our use of the volume fraction as the independent variable in 
the genus analysis is partly motivated by the hope that smoothing on a 
sufficient scale yields a monotonic relationship between mass density in 
the initial conditions and galaxy density at the observed epoch.
Departure from monotonicity in the mass-galaxy relationship, 
non-locality of this relationship, or stochasticity in this mapping 
could violate this assumption, if the smoothing scale is smaller than 
the scale of such effects. Thus, dependence on galaxy color and the 
departures from the Gaussian prediction of the full sample may indicate 
significant complexity in the mass-galaxy relationship on smoothing 
scales $\lambda\sim 5h^{-1}$Mpc.

\section{Conclusions}
\label{sec:conc}

We estimate the two-dimensional genus of an early sample of SDSS
galaxies, which is one of the two largest samples analyzed for 2D topology. 
We find that the genus is consistent with that of a
$\Lambda$CDM simulation and has an approximately Gaussian shape,
although there is a slight meatball shift, consistent with the
simulation. The peaks of the genus curve appear at slightly smaller
absolute $\nu$ values than would be found for a Gaussian distribution
or for the $\Lambda$CDM simulation, but the statistical significance
of this effect is marginal.

We compare the SDSS results with the 2dFGRS genus curve and find that the
two curves have approximately the same shape, when renormalized due to
the different areas of the surveys. 

To test for segregation of galaxies by type, we also split the SDSS
data into thirds using a color criteria and measure the genus curves
of these samples. We find that the genus curve of the reddest galaxies
exhibits a meatball shift, as predicted by N-body/hydrodynamical
simulations (Gott et al. 1996).  However, this effect may sensitively
depend on the number of high-density regions (clusters) in the
sample. The current sample is still relatively small (only 10-20
massive clusters lie within a typical volume of this size) , thus
examination of larger samples is necessary to confirm this
finding. The completed SDSS galaxy redshift survey will include
roughly 10 times more galaxies, allowing us to further test the
detailed dependence of topology on galaxy properties.

\section*{Acknowledgments} 

Funding for the creation and distribution of the SDSS Archive has been
provided by the Alfred P. Sloan Foundation, the Participating
Institutions, the National Aeronautics and Space Administration, the
National Science Foundation, the U.S. Department of Energy, the
Japanese Monbukagakusho, and the Max Planck Society. The SDSS Web site
is http://www.sdss.org/.

The SDSS is managed by the Astrophysical Research Consortium (ARC) for
the Participating Institutions. The Participating Institutions are The
University of Chicago, Fermilab, the Institute for Advanced Study, the
Japan Participation Group, The Johns Hopkins University, Los Alamos
National Laboratory, the Max-Planck-Institute for Astronomy (MPIA),
the Max-Planck-Institute for Astrophysics (MPA), New Mexico State
University, Princeton University, the United States Naval Observatory,
and the University of Washington.  

MSV acknowledges support from NSF grant AST-0071201 and a grant from
the John Templeton Foundation. JRG acknowledges the support from NSF
grant AST-9900772.  We thank Ravi Sheth and Chris Miller for useful
conversations. We thank the anonymous referee for insightful comments.

\end{document}